\documentclass[final,3p,times]{elsarticle}
\usepackage{amssymb}
\usepackage{CJK}
\usepackage{graphics}
\usepackage{amssymb}
\usepackage{amsthm}
\usepackage{setspace}
\usepackage{epsfig}
\usepackage{subfigure}
\usepackage{booktabs}
\usepackage{epstopdf}
\usepackage{empheq}
\usepackage{color}

\biboptions{numbers,sort&compress}

\usepackage{amsmath}
\usepackage{times}
\usepackage{anysize}
\marginsize{2.5cm}{2.5cm}{2cm}{2cm}
\selectfont

 \journal {publication}

\begin{document}
\begin{CJK*}{GBK}{song}
\begin{frontmatter}

\newtheorem{theorem}{Theorem}
\newtheorem{remark}{Remark}
\newtheorem{lemma}{Lemma}
\newtheorem{Proof}{Proof of Theorem.}
\newtheorem{Prf}{Proof of lemma 2.}
\newtheorem{ff}{Proof. }


\title{Progresses on some open problems related to infinitely many symmetries}
\author{S. Y. Lou}
\ead{lousenyue@nbu.edu.cn}
\cortext[cor]{Corresponding author: School of Physical Science and Technology, Ningbo University,
Ningbo, 315211, China}
\address{School of Physical Science and Technology, Ningbo University, Ningbo, 315211, China}

\begin{abstract}	
The quest to reveal the physical essence of the infinitely many symmetries and/or conservation laws that are intrinsic to integrable systems has historically posed a significant challenge at the confluence of physics and mathematics. This scholarly investigation delves into five open problems related to these boundless symmetries within integrable systems by scrutinizing their multi-wave solutions, employing a fresh analytical methodology. For a specified integrable system, there exist various categories of $n$-wave solutions, such as the $n$-soliton solutions, multiple breathers, complexitons, and the $n$-periodic wave solutions (the algebro-geometric solutions with genus $n$), wherein $n$ denotes an arbitrary integer that can potentially approach infinity. Each sub-wave comprising the $n$-wave solution may possess free parameters, including center parameters $c_i$, width parameters (wave number) $k_i$, and periodic parameters (the Riemann parameters) $m_i$. It is evident that these solutions are translation invariant with respect to all these free parameters. We postulate that the entirety of the recognized infinitely many symmetries merely constitute linear combinations of these finite wave parameter translation symmetries. This conjecture appears to hold true for all integrable systems with $n$-wave solutions.
The conjecture intimates that the currently known infinitely many symmetries are not exhaustive, and an indeterminate number of symmetries remain to be discovered. This conjecture further indicates that by imposing an infinite array of symmetry constraints, it becomes feasible to derive exact multi-wave solutions. By considering the renowned Korteweg-de Vries (KdV) equation and the Burgers equation as simple examples, the conjecture is substantiated for the $n$-soliton solutions. It is unequivocal that any linear combination of the wave parameter translation symmetries retains its status as a symmetry associated with the particular solution. This observation suggests that by introducing a ren-variable and a ren-symmetric derivative which serve as generalizations of the Grassmann variable and the super derivative, it may be feasible to unify classical integrable systems, supersymmetric integrable systems, and ren-symmetric integrable systems within a cohesive hierarchical framework. Notably, a ren-symmetric integrable Burgers hierarchy is explicitly derived. Both the supersymmetric and the classical integrable hierarchies are encompassed within the ren-symmetric integrable hierarchy.
\end{abstract}

\end{frontmatter}

% \linenumbers

 \section{Introduction}

Over five decades ago, the pioneering work of Miura, Gardner, and Kruskal unveiled the potential for an infinitely many conservation laws within the Korteweg-de Vries (KdV) equation, a revelation that has since profoundly influenced the study of nonlinear dynamics \cite{KdV1}. Subsequent investigation by Olver has identified recursion operators within a class of (1+1)-dimensional integrable systems, including not only the KdV but also the modified KdV (MKdV) and sine-Gordon (sG) equations, facilitating the recursive derivation of an infinitely many symmetries and conservation laws \cite{Olver}. The contributions of Fuchssteiner highlighted the hereditary properties of these recursion operators, suggesting a universality across all equations within an integrable hierarchy \cite{HereD}. In the (2+1)-dimensional context, the master-symmetry method and formal series symmetry approach have been instrumental in uncovering an abundance of symmetries \cite{Master, PRLLou, HaoL, HLL}.

Despite the extensive applications of these symmetries in addressing nonlinear physical phenomena, several fundamental questions persist, prompting a reexamination of their physical underpinnings and mathematical integrity.\\
\bf Open Problem 1: \it Physical explanations of infinitely many symmetries and conservation laws
\\
\rm
The elucidation of the physical interpretations of the infinitely many symmetries and conservation laws inherent to integrable systems represents a profound and enduring question at the intersection of physics and mathematics. While it is widely acknowledged that integrable systems are characterized by an infinitely many symmetries and conservation laws, only a select few possess explicit, tangible physical interpretations, such as those related to space-time translational invariance, Galilean or Lorentz invariance, scaling invariance, as well as the conservation of mass, momentum, and energy.\\
\bf Open Problem 2: \it Completeness of known infinitely many symmetries
\\
\rm
For nonlinear physical systems with finite $n$ freedoms described by an ordinary differential equation (ODE) system, it suffices to identify a finite number of symmetries or conserved quantities, as the general solution of the model possesses a finite number of arbitrary constants. However, for systems with infinitely many freedoms described by partial differential equation (PDE) systems, assessing the completeness of their symmetries is exceedingly challenging. This paper will conclude that the known infinitely many symmetries for all integrable systems are far from complete.\\
\bf Open Problem 3: \it Finding multi-wave solutions via generalized symmetries
\\
\rm
While symmetries and conservation laws are instrumental in deriving general solutions for integrable ODE systems, the analogous application within the context of integrable PDEs presents a significant challenge. Despite the fact that several Lie point symmetries have been successfully applied to identify single periodic cnoidal wave solutions, including solitary waves, and Painlev\'e solutions, the utilization of the vast array of general symmetries to find exact solutions in PDE systems is an unresolved issue. The symmetry constraint method, which leverages nonlocal symmetries, has been effective in discovering algebro-geometric solutions \cite{Cao}, and these nonlocal symmetries have also facilitated the derivation of Darboux transformations \cite{LouLi,TangL}. In this paper, we propose a novel approach to uncover multi-wave solutions, particularly focusing on the multi-soliton solutions, by harnessing the power of generalized local symmetries.\\
\bf Open problem 4: \it Loss of certain orders of generalized symmetries
\\
\rm
The existence of an infinitely many symmetries with varying orders in integrable systems is well-documented. However, for certain key integrable systems, specific orders of generalized symmetries appear to be absent. For instance, the Sawada-Kortera (SK) \cite{SK} and Kaup-Kupershmidt (KK) \cite{KK} equations exhibit generalized symmetries of orders $6n+1$ and $6n+5$ ($n=0,\ 1,\ \ldots,\ \infty$), yet the $6n+3$ order symmetries are missing. Furthermore, for well-known equations like KdV and MKdV, there are infinitely many generalized symmetries of odd orders. This raises the question of the existence of even order symmetries and, more broadly, the possibility of fractional order symmetries. This paper addresses this open problem by introducing Grassmann variables and Ren variables \cite{Ren, Ren1}.\\
\bf Open Problem 5: \it Arbitrary constants in special solutions via symmetry transformations
\\
\rm
It is a well-established principle that for a particular solution of a given system, denoted as $u_0(x, t)$, the incorporation of an arbitrary constant is feasible through the application of a symmetry. Specifically, by invoking the invariance under spatial translation in the $x$-direction and temporal translation in the $t$-direction, it is possible to seamlessly integrate two arbitrary constants, represented as $\{x_0, t_0\}$. This results in a transformation of the solution to $u_0(x-x_0, t-t_0)$, which retains its validity within the framework of the original model.
The question that arises is whether the leverage of an infinitely many symmetries could permit the introduction of an infinite series of arbitrary constants into a specific solution.\\
To delve into these open problems, it is imperative to initiate the investigation with a focus on certain classes of special solutions. A plethora of exact solutions with significant physical implications has been identified, encompassing multi-soliton solutions, algebro-geometric solutions, and similarity solutions.  The concept of solitons, initially observed by Russell in 1834 \cite{1834} and subsequently named by Zabusky and Kruskal in 1965 \cite{ZK}, has permeated various physical disciplines. These solitons are encapsulated within the framework of integrable systems, which include, but are not limited to, the KdV equation \cite{KdV,Bq,KdV1,KdV2}, the MKdV equation \cite{MKdV}, the sG equation \cite{SG,SG1}, the nonlinear Schr\"odinger (NLS) equation \cite{NLS}, the Boussinesq equation \cite{Bq}, and the Burgers equation \cite{BE}.
\\
In the forthcoming Sections 2 and 3 of this paper, taking the multi-soliton solutions of the KdV equation and Burgers equation as simple examples, the physical interpretations of the infinitely many $K$-symmetries and $\tau$-symmetries are explicitly pointed out. It is revealed that for the $n$-soliton solution of the KdV equation and the $n$-resonant soliton solutions of the Burgers equation, the infinite many $K$-symmetries are merely linear combinations of the center translation invariance pertaining to each soliton. Similarly, the infinite many $\tau$-symmetries are identified as linear combinations of the soliton width translation invariance and the soliton center translation invariance.
Advancing to Section 4, we delve into the non-completeness of these infinitely many symmetries,  particularly within the milieu of the single soliton solution of the KdV equation. The revelations from Sections 2 and 3 impart a provocative insight for (1+1)-dimensional integrable systems: the postulation that if $\Phi$ is identified as a recursion operator, then its arbitrary $\alpha$-th root may also encapsulate the properties of a recursion operator. This proposition intimates the tantalizing possibility of the existence of generalized symmetries across the continuum of orders, transcending the conventional integer bounds and potentially encompassing fractional orders.
Continuing in Section 5, we extend our inquiry to the exploration of the $\alpha$-th roots of the conventional recursion operators, leveraging the conception of ren variables, which represents an innovative generalization of the traditional Grassmann variables \cite{Ren,Ren1}. Notably, for the class of C-integrable systems \cite{Calo}, such as the Burgers equation, the explicit derivation of these $\alpha$-th roots of the recursion operators is within our grasp. This discovery paves the way for an integrated hierarchical schema that seamlessly incorporates supersymmetric, ren-symmetric, and classical integrable systems.
Section 6 heralds the introduction of a symmetry conjecture aimed at the multi-wave solution of integrable systems. We conjecture that the extant infinitely many $K$-symmetries and $\tau$-symmetries are fundamentally linearized combinations of finite wave parameter translation invariance.
Building upon this conjecture, Section 7 delineates an innovative methodology for resolving multi-wave solutions, with a pronounced emphasis on multi-soliton solutions. The last section concludes with a synthesis of our findings and an in-depth discussion on the implications of infinitely many symmetries.

\section{ Physical interpretations of the infinitely numerous symmetries inherent in the KdV equation}
To elucidate the concept, we anchor our discussion in the celebrated KdV equation,
\begin{equation}\label{KdV}
u_t=u_{xxx}+6uu_x,
\end{equation}
a paradigmatic model in the study of shallow water waves and soliton theory.
The KdV equation was first introduced by Boussinesq \cite{Bq} and rediscovered by Diederik Korteweg and Gustay de Vries \cite{KdV}. The KdV equation has various connections to physical problems. It approximately describes the evolution of long, one-dimensional waves in many physical settings, including shallow-water waves with weakly non-linear restoring forces, long internal waves in a density-stratified ocean, ion acoustic waves in a plasma, acoustic waves on a crystal lattice \cite{KdV2} and the two-dimensional quantum gravity \cite{Guo}, The KdV equation can be solved using the inverse scattering transform and other methods such as those applied to other integrable systems \cite{GGKM}.

A symmetry, denoted by $\sigma$, of the KdV equation is characterized as a solution to its linearized equation,
\begin{equation}\label{symkdv}
\sigma_t=\sigma_{xxx}+6(u\sigma)_x.
\end{equation}
This definition encapsulates the notion that the KdV equation maintains its form under the infinitesimal transformation $u \rightarrow u+\epsilon \sigma$, where $\epsilon$ is an infinitesimal parameter.

The KdV equation is renowned for its infinite sets of symmetries, commonly referred to as the $K$-symmetries and $\tau$-symmetries, which are articulated as follows,
\begin{eqnarray}
K_{n+1}&=&\Phi^nu_x,\ \Phi\equiv \partial_x^2+4u+2u_x\partial_x^{-1},\label{K}\\
\tau_{n+1}&=&3tK_{n+1}+\Phi^n \frac12 =\Phi^n\left(3tu_x+\frac12\right),\ n=0,\ 1,\ ...,\ \infty,\label{tau}
\end{eqnarray}
where the operator $\Phi$ shown in \eqref{K} is just the known recursion operator of the KdV equation \cite{Olver}.
The first quartet of $K$-symmetries are explicitly given by
\begin{eqnarray}
K_1&=&u_x,\nonumber\\
K_2&=&\big(u_{xx}+3u^2\big)_x=u_t,\nonumber\\
K_3&=&\big(u_{xxxx}+10uu_{xx}+5u_x^2+10u^3\big)_x,\nonumber\\
K_4&=&\big[u_{xxxxxx}+14uu_{xxxx}+28u_xu_{xxx}+21u_{xx}^2+70u(uu_{x})_x+35u^4\big]_x, \label{K4}
\end{eqnarray}
and the initial subset of $\tau$-symmetries is delineated as
\begin{eqnarray}
\tau_1&=&3tu_x+\frac12=3tK_1+\frac12,\nonumber\\
\tau_2&=&3tu_t+xu_x+2u=3tK_2+xK_1+2u,\nonumber\\
\tau_3&=&3tK_3+xK_2+2K_1\int u\mbox{\rm d}x+4u_{xx}+8u^2,\nonumber\\
\tau_4&=&3tK_4+xK_3+2K_2\int u\mbox{\rm d}x +6K_1\int u^2\mbox{\rm d}x+6K_{2x}+12uu_{xx}+32u^3.\label{T4}
\end{eqnarray}

	To date, only the initial members of these symmetry sets, $\{K_1,\ K_2\}$ and $\{\tau_1,\ \tau_2\}$, have been ascribed concrete physical interpretations. Specifically, $K_1$ is associated with spatial translational invariance, $K_2$ corresponds to temporal translational invariance, $\tau_1$ encapsulates the invariance under Galilean transformation, and $\tau_2$ is responsible for scaling transformation invariance. The physical interpretations of the remaining symmetries remain an enigma.
	
	Inextricably linked to the $K$-symmetries, there emerges an infinite number of conservation laws,
	\begin{eqnarray}
	\rho_{n+1,t}=J_{n+1,x},\label{CL}
	\end{eqnarray}
	with the conserved densities,
	\begin{equation}
	\rho_{n+1}=\partial_x^{-1}K_{n+1}, \label{Krho}
	\end{equation}
	which are expressible through
	\begin{eqnarray}
	\rho_{n+1}=\tilde{\Phi}^n u,\qquad \tilde{\Phi}\equiv \partial_x^{-1}\Phi\partial_x
	= \partial_x^2+4u-2\partial_x^{-1} u_x. \label{rho}
	\end{eqnarray}
Upon the exclusion of trivially conserved total derivatives along the spatial coordinate $x$ and the normalization by inconsequential scalar factors, the explicit expressions of the quintessential conserved densities $\rho_{n+1}$ are rendered in their simplified forms as follows,
\begin{align*}
&\rho_1 = u, \quad &\text{(mass density)} \\
&\rho_2 = u^2, \quad &\text{(momentum density)} \\
&\rho_3 = u^3 - \frac{1}{2}u_x^2, \quad &\text{(energy density)} \\
&\rho_4 = u^4 - 2uu_x^2 + \frac{1}{5}u_{xx}^2, \\
&\rho_5 = u^5 - 5u^2u_x^2 + uu_{xx}^2 - \frac{1}{14}u_{xxx}^2.
\end{align*}
When the KdV equation serves as a model for the description of shallow water wave phenomena, it is recognized that only the inaugural trio of conserved densities  $\{\rho_1,\ \rho_2,\ \rho_3\}$ have been endowed with established physical interpretations. These correspond to the conservation laws of mass, momentum, and energy, respectively, within the hydrodynamical context \cite{cls}. The subsequent conserved densities, $\rho_4$ and $\rho_5$, while mathematically derived, await elucidation of their physical significance in the context of fluid dynamics.

For a specified solution $ u = u_0(x,\ t)$  of the KdV equation \eqref{KdV}, the employment of four specific symmetries, namely $K_1,\ K_2,\ \tau_1$ and $\tau_2$, permits the incorporation of four arbitrary constants $x_0,\ t_0,\ g$ and $s$. This leads to the construction of a novel solution expressed by the transformation,
$$u=k^2u_0(k(x-gt-x_0),\ k^3(t-t_0))-\frac{g}{6},$$
where the scaling parameter has been renamed as $k=\exp(-s\epsilon)$ and the $x$-translation parameter has been re-denoted as $x_0-gt_0\rightarrow x_0$.

The quest to introduce an augmented set of arbitrary constants via the utilization of additional $K$-symmetries and $\tau$-symmetries for the given solution $u=u_0(x,\ t)$ remains an open question in the field.

To elucidate the potential physical interpretations, we shall specify the exact form of the solution $u_0(x,\ t)$ for the KdV equation. The $n$-soliton solution stands out as a fundamental and significant archetype, embodying the structure,
\begin{equation}\label{soln}
u_{ns}=2\left(\ln F\right)_{xx},
\end{equation}
where $F$ is given by the summation over all permutations of $\mu =\{\mu_i\ | \ \ i=1,\ 2,\ \ldots,\ n\}, \ \mu_{i}=0,\ 1,\ \forall\ i$, with
\begin{eqnarray}
F =\sum_{\mu}\exp\left(\sum_{j=1}^n\mu_j\xi_j+\sum_{1\leq j<l}^n\mu_j\mu_l\theta_{jl}\right),
\end{eqnarray}
\begin{equation}\label{xi}
\xi_j=k_jx+k_j^3t+c_j, \quad \exp(\theta_{jl})=\left(\frac{k_j-k_l}{k_j+k_l}\right)^2.
\end{equation}

The one-, two-, and three-soliton solutions are special cases with the following simplified expressions,
\begin{equation}\label{sol1}
u_{1s}=2p_1^2\mbox{\rm sech}^2(\eta_1),\ \eta_i=\frac12\xi_i=p_ix+4p_i^3t+b_i,\ p_i=\frac12k_i,\
b_i=\frac12c_i,
\end{equation}
\begin{equation}\label{sol2}
u_{2s}=\frac{(k_1^2-k_2^2)[k_1^2-k_2^2+k_1^2\cosh(2\eta_2)+k_2^2\cosh(2\eta_1)]}
{[k_1\cosh(\eta_1)\cosh(\eta_2)-k_2\sinh(\eta_1)\sinh(\eta_2)]^2},
\end{equation}
and
\begin{eqnarray}\label{sol3}
u_{3s}&=&2(\ln F_{3s})_{xx},\nonumber\\
F_{3s}&=&1+\mbox{\rm e}^{\xi_1}+\mbox{\rm e}^{\xi_2}+\mbox{\rm e}^{\xi_3}+a_{12}\mbox{\rm e}^{\xi_1+\xi_2}
+a_{13}\mbox{\rm e}^{\xi_1+\xi_3}+a_{23}\mbox{\rm e}^{\xi_2+\xi_3}
+a_{12}a_{13}a_{23}\mbox{\rm e}^{\xi_1+\xi_2+\xi_3}.
\end{eqnarray}

For the specified solution $u_0 = u_{ns}$, it becomes evident that there are $2n$ physically significant symmetries, given by
\begin{eqnarray}
\sigma_{c_m}\equiv u_{c_m}=2\left(\frac{F_{c_m}}{F}\right)_{xx},\quad F_{c_m}
=\sum_{\mu}\mu_m\exp\left(\sum_{j=1}^n\mu_j\xi_j
+\sum_{1\leq j<l}^n\mu_j\mu_l\theta_{jl}\right),\ m=1,\ 2,\ \ldots,\ n,\label{cm}
\end{eqnarray}
and
\begin{eqnarray}
\sigma_{k_m}\equiv u_{k_m}=2\left(\frac{F_{k_m}}{F}\right)_{xx},\quad
F_{k_m}=\sum_{\mu}\left(\mu_m\xi_{m,k_m}
+\sum_{1\leq i<p}^n\mu_i\mu_p\theta_{ip,k_m}\right)\exp\left(\sum_{j=1}^n\mu_j\xi_j
+\sum_{1\leq j<l}^n\mu_j\mu_l\theta_{jl}\right),\label{km}
\end{eqnarray}
respectively. The physical interpretations of the symmetries $\sigma_{c_m}$ and $\sigma_{k_m}$ are self-evident. $\sigma_{c_m}$ pertains to the center translation invariance of the $m$-th soliton, whereas $\sigma_{k_m}$ corresponds to the wave number translation invariance of the $m$-th soliton,
with the wave number $k_m$ being associated with the amplitude, width, and velocity of the $m$-th soliton.

The natural and pertinent inquiry is whether there exist any relations among the symmetries $\{K_{n+1}, \tau_{n+1}\}$ and $\{\sigma_{c_m}, \sigma_{k_m}\}$?

For the $K$-symmetries \eqref{K} with $u=u_{ns}$ specified in \eqref{soln}, it is readily discernible that
\begin{eqnarray}
\left. K_1\right|_{u=u_{ns}}&=&\left.u_x\right|_{u=u_{ns}}
=\sum_{m=1}^n k_m u_{ns,c_m}=\sum_{m=1}^n k_m\sigma_{c_m}, \nonumber\\
\left. K_2\right|_{u=u_{ns}}&=&\left.\Phi u_x\right|_{u=u_{ns}}
=\sum_{m=1}^n k_m^{3} u_{ns,c_m}=\sum_{m=1}^n k_m^{3}\sigma_{c_m},\nonumber\\
\left. K_3\right|_{u=u_{ns}}&=&\left.\Phi^{2}u_x\right|_{u=u_{ns}}
=\sum_{m=1}^n k_m^{5} u_{ns,c_m}=\sum_{m=1}^n k_m^{5}\sigma_{c_m},\nonumber\\
 &  \vdots &\nonumber\\
\left. K_i\right|_{u=u_{ns}}&=&\left.\Phi^{i-1}u_x\right|_{u=u_{ns}}
=\sum_{m=1}^n u_{ns,c_m}k_m^{2i-1}=\sum_{m=1}^n k_m^{2i-1}\sigma_{c_m},\
i=3,\ 4,\ \ldots, \ \infty. \label{Kicm}
\end{eqnarray}
From the relation \eqref{Kicm}, we ascertain that for the given solution \eqref{soln}, the infinitely many $K$-symmetries are not autonomous, only $n$ of them, namely, $K_i,\ i=1,\ \ldots, \ n$, are independent while the remaining symmetries can be linearly constructed by them,
\begin{eqnarray}
&&\left. K_i\right|_{u=u_{ns}}=\left.\Phi^{i-1}u_x\right|_{u=u_{ns}}
=\left.\sum_{m=1}^n k_m^{2i-1}\frac{\Delta_{nm}}{\Delta_n}\right|_{u=u_{ns}},\
i=n+1,\ n+2,\ \ldots, \ \infty, \label{Kcm}
\end{eqnarray}
where $\Delta_n$ denotes the determinant of the $n\times n$ matrix $M$, i.e., $\Delta_n\equiv{\det(M)}$ with
\begin{eqnarray}
 M=\left(\begin{array}{ccccccc}
 k_1 & k_2 & \cdots & k_m & \cdots & k_{n-1} & k_n\\
 k_1^3 & k_2^3 & \cdots &k_m^3 & \cdots &  k_{n-1}^3 & k_n^3\\
 \vdots & \vdots & \vdots &\vdots & \vdots &  \vdots & \vdots \\
 k_1^{2n-1} & k_2^{2n-1} & \cdots &k_m^{2n-1} & \cdots &  k_{n-1}^{2n-1} & k_n^{2n-1} \end{array}
 \right),\label{M}
\end{eqnarray}
and $\Delta_{nm}$ represents the determinant of the $n\times n$ matrix $M_m$, $\Delta_{nm}\equiv{\det(M_m)}$,
\begin{eqnarray}
 M_m=\left(\begin{array}{ccccccc}
 k_1 & k_2 & \cdots & K_1 & \cdots & k_{n-1} & k_n\\
 k_1^3 & k_2^3 & \cdots &K_2 & \cdots &  k_{n-1}^3 & k_n^3\\
 \vdots & \vdots & \vdots &\vdots & \vdots &  \vdots & \vdots \\
 k_1^{2n-1} & k_2^{2n-1} & \cdots &K_n & \cdots &  k_{n-1}^{2n-1} & k_n^{2n-1} \end{array}
 \right).\label{Mm}
\end{eqnarray}
For instance, for the two-soliton solution, $u_{ns}=u_{2s}$ given by \eqref{sol2}, there are merely two independent $K$-symmetries (say, $\left. K_1\right|_{u=u_{2s}}$ and $\left. K_2\right|_{u=u_{2s}}$), while the rest are linear combinations of them as stipulated by \eqref{Kcm}, i.e.,
\begin{eqnarray}
&&K_{m\geq3}=k_1^2k_2^2(A_{1m} K_1+A_{2m}K_2),\ A_{1m}=\frac{k_2^{2m-4}-k_1^{2m-4}}{k_1^2-k_2^2},\
A_{2m}=\frac{k_1^{2m-2}-k_2^{2m-2}}{k_1^2k_2^2(k_1^2-k_2^2)},
\label{Kc2}
\end{eqnarray}
where, for notational simplicity, $\left. K_i\right|_{u=u_{2s}}, \ i=1,\ 2,\ \ldots, $ have been consistently denoted as $K_i$.

It is rational to posit that the $\tau$-symmetries \eqref{tau} with $u=u_{ns}$ specified in \eqref{soln} constitute linear combinations of the $k_i$-translation invariance $\sigma_{k_i}$ and the $c_i$-translation invariance $\sigma_{c_i}$. Using the two-soliton solution $u_{2s}$ \eqref{sol2} as an illustration, we obtain,
\begin{eqnarray}
\tau_2&=&k_1\sigma_{k_1}+k_2\sigma_{k_2},\nonumber\\
\tau_3&=&k_1^3\sigma_{k_1}+k_2^3\sigma_{k_2}-2k_1(k_1+2k_2)\sigma_{c_1}-2k_2(2k_1+k_2)\sigma_{c_2},\nonumber\\
\tau_4&=&k_1^5\sigma_{k_1}+k_2^5\sigma_{k_2}-4k_1(k_1^3+k_1^2k_2+k_2^3)\sigma_{c_1}
-4k_2(k_1^3+k_1k_2^2+k_2^3)\sigma_{c_2}
\nonumber\\
&=&-k_1^2k_2^2\tau_2+(k_1^2+k_2^2)\tau_3+2(k_1+k_2)(k_1k_2 K_1-K_2).\label{t4k4}
\end{eqnarray}
The interrelations among other $\tau$-symmetries and $\{\tau_2,\ \tau_3,\ K_1,\ K_2\}$ or, equivalently, $\{\sigma_{k_1},\ \sigma_{k_2},\ \sigma_{c_1},\ \sigma_{c_2}\}$ can be recursively determined using $\tau_{n+1}=\Phi \tau_n,\ K_{n+1}=\Phi K_n$, and the expressions of $K_3$ \eqref{Kc2} and $\tau_4$ \eqref{t4k4}. The ultimate general results may be encapsulated as
\begin{eqnarray}
\tau_{m\geq4}&=&[(k_1^2+k_2^2)A_{1m}+k_1^2k_2^2A_{2m}]\tau_2-A_{1m}\tau_3+\frac{2B_{1m}}{k_1k_2(k_1^2-k_2^2)}K_1
+\frac{2B_{2m}}{k_1k_2(k_1+k_2)}K_2,\label{tm}
\end{eqnarray}
where
\begin{eqnarray}
B_{1m}&=&(1-m)[(k_1^5-k_2^5)A_{1m}+k_1^2k_2^2(k_1^3-k_2^3)A_{2m}]+(k_1^3+k_2^3)(k_1^2-k_2^2)A_{1m}
+k_1^2k_2^2(k_1^3-k_2^3)A_{2m},\nonumber\\
B_{2m}&=&(m-1)[(k_1^2+k_1k_2+k_2^2)A_{1m}+k_1^2k_2^2A_{2m}]
-(k_1+k_2)^2A_{1m}-k_1^2k_2^2A_{2m},
\end{eqnarray}
while $A_{1m}$ and $A_{2m}$ are defined in \eqref{Kc2}.

It merits mention that the $\tau_1$ symmetry pertains to the Galileo transformation invariance and does not correlate with the symmetries $\sigma_{k_i}$, which relate to the amplitude, width, and velocity of the $i$-th soliton.

\section{Physical interpretations of the infinitely numerous symmetries inherent in the Burgers equation}
To elucidate the universality of the physical interpretations associated with the infinitely many $K$-symmetries and $\tau$-symmetries, we consider the simpler integrable model, the Burgers equation
\begin{equation}
v_t=v_{xx}+2vv_x \label{BE}
\end{equation}
as an illustrative example. It is recognized that the Burgers equation possesses a recursion operator $\Phi_b$,
\begin{equation}
\Phi_b=\partial_x+v+v_x\partial_x^{-1}. \label{phib}
\end{equation}
The $K$-symmetries and the $\tau$-symmetries of the Burgers equation \eqref{BE} are characterized by
\begin{eqnarray}
K_{m+1}&=&\Phi_b^{m}v_x,\ m=0,\ 1,\ \ldots,\ \infty, \label{KBE}\\
\tau_{m+1}&=&\Phi_b^{m}(2tv_x+1),\ m=0,\ 1,\ \ldots,\ \infty. \label{TBE}
\end{eqnarray}
The $n$ resonant soliton solutions of the Burgers equation \eqref{BE} adopt the form
\begin{equation}
v=v_{ns}=\left(\ln F_{ns}\right)_x,\ F_{ns}=1+\sum_{i=1}^n\exp(k_ix+k_i^2t+c_i). \label{vns}
\end{equation}
For these $n$ resonant soliton solutions \eqref{vns}, there exist $2n$ subwave parameter translation symmetries,
\begin{eqnarray}
\sigma'_{c_j}&=&\partial_{c_j}v_{ns}=v_{ns,c_j}=\left(F_{ns}^{-1}F_{ns,c_j}\right)_x,\
F_{ns,c_j}=\exp(k_jx+k_j^2t+c_j),\ j=1,\ 2,\ \ldots,\ n, \label{vnscj}
\nonumber\\
\sigma'_{k_j}&=&\partial_{k_j}v_{ns}=v_{ns,k_j}=\left((F_{ns}^{-1}F_{ns,k_j}\right)_x,\
F_{ns,k_j}=(x+2k_jt)\exp(k_jx+k_j^2t+c_j),\ j=1,\ 2,\ \ldots,\ n. \label{vnskj}
\end{eqnarray}
Analogous to the KdV case, the infinitely many $K$-symmetries \eqref{KBE} are merely linear combinations of the $n$ center translation symmetries \eqref{vnscj}, albeit with $n$ potentially approaching infinity,
\begin{eqnarray}
\left.K_{m+1}\right|_{v=v_{ns}}&=&\sum_{i=1}^nk_i^{m+1}\sigma'_{c_i},\ m=0,\ 1,\ \ldots,\ \infty, \label{KBEscj}
\end{eqnarray}
and the infinitely many $\tau$-symmetries \eqref{TBE} are solely linear combinations of the $n$ wave number translation symmetries and the $n$ center translation symmetries \eqref{vnscj}
\begin{eqnarray}
\left.\tau_{m+1}\right|_{v=v_{ns}}&=&\sum_{i=1}^nk_i^{m}\sigma'_{k_i}
+(m-1)\sum_{i=1}^nk_i^{m-1}\sigma'_{c_i},\ m=1,\ 2,\ \ldots,\ \infty. \label{KBEskj}
\end{eqnarray}
In equation \eqref{KBEskj}, the inclusion of $m=0$ is precluded due to the correspondence of $\tau_1$ with Galilean invariance, an attribute that bears no relation to wave number translational invariance. This implies that for the fixed, exact $n$-resonant soliton solution \eqref{vns}, the myriad $K$-symmetries and $\tau$-symmetries (with the exception of $\tau_1$), as delineated by \eqref{KBE} and \eqref{TBE}, do not exhibit independence, solely $2n$ of these symmetries (specifically, $\{\tau_{m+1},\ K_{m}\ |\ m=1,\ 2,\ \ldots,\ n\}$ are autonomous, while all others manifest as linear admixtures thereof. This relationship is articulated by
\begin{eqnarray}
\left.K_{i>n}\right|_{v=v_{ns}}&=&\left.\Phi_b^{i-1}v_x\right|_{v=v_{ns}}
=\sum_{m=1}^nk_m^i\frac{\Gamma_{m}}{\Gamma}, \label{KKn}\\
\left.\tau_{i>n+1}\right|_{v=v_{ns}}&=&\sum_{m=1}^nk_m^i\frac{\Gamma'_{m}}{\Gamma}
+(i-2)\sum_{m=1}^nk_m^{i-2}\frac{\Gamma_{m}}{\Gamma}, \label{TTn}
\end{eqnarray}
wherein $\Gamma$ signifies the determinant of the square matrix $B$ of order $n$, characterized by
\begin{eqnarray}
 B=\left(\begin{array}{ccccccc}
 k_1 & k_2 & \cdots & k_m & \cdots & k_{n-1} & k_n\\
 k_1^2 & k_2^2 & \cdots &k_m^2 & \cdots &  k_{n-1}^2 & k_n^2\\
 \vdots & \vdots & \vdots &\vdots & \vdots &  \vdots & \vdots \\
 k_1^{n} & k_2^{n} & \cdots &k_m^{n} & \cdots &  k_{n-1}^{n} & k_n^{n} \end{array}
 \right).\label{B}
\end{eqnarray}
The term $\Gamma_{m}$ denotes the determinant of the matrix $B_m$ of equivalent dimensions, defined by
\begin{eqnarray}
 B_m=\left(\begin{array}{ccccccc}
 k_1 & k_2 & \cdots & K_1 & \cdots & k_{n-1} & k_n\\
 k_1^2 & k_2^2 & \cdots &K_2 & \cdots &  k_{n-1}^2 & k_n^2\\
 \vdots & \vdots & \vdots &\vdots & \vdots &  \vdots & \vdots \\
 k_1^{n} & k_2^{n} & \cdots &K_n & \cdots &  k_{n-1}^{n} & k_n^{n} \end{array}
 \right).\label{Bm}
\end{eqnarray}
Conversely, $\Gamma'_m$ symbolizes the determinant of the matrix $B'_m$ defined by
\begin{eqnarray}
 B'_m=\left(\begin{array}{ccccccc}
 k_1 & k_2 & \cdots & \tau_2 & \cdots & k_{n-1} & k_n\\
 k_1^2 & k_2^2 & \cdots &\tau_3 & \cdots &  k_{n-1}^2 & k_n^2\\
 \vdots & \vdots & \vdots &\vdots & \vdots &  \vdots & \vdots \\
 k_1^{n} & k_2^{n} & \cdots &\tau_{n+1} & \cdots &  k_{n-1}^{n} & k_n^{n} \end{array}
 \right).\label{Bm'}
\end{eqnarray}
Upon constraining the system size parameter to $n=2$, equations \eqref{KKn} and \eqref{TTn} can be condensed into more compact forms, as depicted by
\begin{eqnarray}
\left.K_{i\geq3}\right|_{v=v_{2s}}&=&-k_1k_2C_{i2}K_1
+C_{i1}K_2,\ C_{im}=\sum_{j=0}^{i-m-1}k_1^jk_2^{i-j-m-1},\ m=1,\ 2,\ 3,\ 4, \label{KK2}\\
\left.\tau_{i>3}\right|_{v=v_{2s}}&=&-k_1k_2C_{i2}\tau_2
+C_{i1}\tau_3-\left[C_{i2}
+(i-2)k_1k_2C_{i4}\right]K_1
+(i-2)C_{i3}K_2. \label{TT2}
\end{eqnarray}

\section{Incompleteness of the infinitely many known symmetries of the potential KdV equation}
In the preceding analytical discourse, it is elucidated that within the confines of a fixed $n$-soliton solution for a (1+1)-dimensional integrable system, the ostensibly infinite $K$-symmetries and $\tau$-symmetries are effectively degenerated to a finite set comprising $2n$ symmetries. Nonetheless, for any given special solution, the symmetry equation retains the capacity to encapsulate infinitely many symmetries. For instance, the potential KdV (PKdV) equation, articulated as
\begin{equation}
p_t=p_{xxx}+3p_x^2, \label{pkdv}
\end{equation}
which is intricately linked to the KdV equation through the differential relationship $u=p_x$, exhibits a symmetry equation for the particular solution $p=p_0$ that assumes the following form
\begin{equation}
\sigma_t=\sigma_{xxx}+6p_{0x}\sigma_x. \label{s0}
\end{equation}
It is self-evident that the symmetry equation \eqref{s0} remains a PDE endowed with an inexhaustible multitude of degrees of freedom. Paralleling the attributes of the KdV equation, for the $n$-soliton solutions of the PKdV equation, the extant infinite $K$-symmetries and $\tau$-symmetries are revealed to be non-exhaustive, with only the inaugural $n$ $K$-symmetries and $n+1$ $\tau$-symmetries manifesting independence, the remainder emerging as linear combinations of these foundational symmetries.
\\
In this section, we engage in an examination of the symmetries of the PKdV equation, predicated on the fixed single soliton solution
\begin{equation}
p_0=2k\tanh(kx+4k^3t+c)\label{p0}
\end{equation}
 with two arbitrary constants $k$ and $c$. In congruence with the analytical findings of Section 2, for the aforesaid one-soliton solution, the purported infinite $K$-symmetries and $\tau$-symmetries are degenerated to finite ones, delineated as
\begin{equation}
K_i=(2k)^{2i-2}K_1,\ i=1,\ 2,\ \ldots,\ \infty. \label{Kip0}
\end{equation}
Yet, the symmetry equation \eqref{s0}, when coupled with the single soliton solution \eqref{p0}, continues to harbor an infinitely many solutions. Implementing a transformation of the independent variables
\begin{equation}\label{y}
\{\tanh(kx+4k^3t+c),\ k^3 t\}\rightarrow \{y,\ \tau \},
\end{equation}
the symmetry equation \eqref{s0} is transmuted into a novel form
\begin{equation}
\sigma_{\tau}+(y^2-1)^3\sigma_{yyy}+6y(1-y^2)^2\sigma_{yy}-6(1-y^2)^2\sigma_y=0.\label{sy}
\end{equation}
The resolution of this transformed symmetry equation \eqref{sy} is achieved through the conventional variable separation approach, yielding a series representation
\begin{equation}
\sigma=\sum_{i=1}^\infty a_i \sigma_i,\ \sigma_i=\exp(\lambda_i \tau)Y_i(y), \label{rsi}
\end{equation}
where $\{a_i, \lambda_i\}$ are arbitrary constants and $Y_i\equiv Y_i(y)$ is determined by a specific differential equation
\begin{equation}
(y^2-1)^3Y_{iyyy}+6y(1-y^2)^2Y_{iyy}-6(1-y^2)^2Y_{iy}+\lambda_iY_i=0.\label{Yy}
\end{equation}
The general solution for $Y_i(y)$ is constructed as an integral involving exponential functions
\begin{equation}
Y_i=\sum_{j=1}^3d_j \exp\left(\int \frac{16(1-y^2)^2-4b_jy^2+8b_j+b_j^2
}{(1-y^2)(\lambda_i+8y-8y^3)}\mbox{\rm dy} \right),\label{Yi}
\end{equation}
with $d_j$ as arbitrary constants and $b_j$, for $j = 1, 2, 3$, being the roots of a cubic equation
\begin{equation}
b^3+8b^2+16 b+\lambda_i^2=0.\label{bj}
\end{equation}
Consequently, an infinitely many novel symmetries for the PKdV equation \eqref{pkdv} is revealed
\begin{equation}
\sigma_i=\exp\left[\lambda_i k^3t+\int^{\tanh(kx + 4k^3t + c)} \frac{16(1-y^2)^2-4b_iy^2+8b_i+b_i^2
}{(1-y^2)(\lambda_i+8y-8y^3)}\mbox{\rm dy}\right],\ i=1,\ 2,\ \ldots,\ \infty.\label{sj}
\end{equation}
Here, $b_i$ is correlated with arbitrary $\lambda_i$ through the equation $b_i^3 + 8b_i^2 + 16b_i + \lambda_i^2 = 0$.
\\
It transpires through rigorous analysis that the conventional $K$-symmetries, as encapsulated by equation \eqref{Kip0}, bear relevance exclusively to the scenario where the $\lambda_i$ equals zero in relation to the symmetries expressed by equation \eqref{sj}. Conversely, the symmetries delineated by \eqref{sj} for $\lambda_i \neq 0$ are identified as a novel set of symmetries that are hitherto unexplored, and are intrinsically associated with the solitary wave solution characterized by equation \eqref{p0}.

\section{Supersymmetric and ren-symmetric integrable Burgers hierarchies}

From the formulations \eqref{Kicm} pertinent to the KdV equation, it is discerned that when the operator $\Phi^i$ is applied to the fundamental symmetry $u_x$, with $i$ being a positive integer, for the $n$-soliton solutions, the sole modifications that transpire are the substitutions of the combination coefficients, specifically $k_m \rightarrow k_m^{2i+1}$. It becomes evidently clear that if the index $i$ in \eqref{Kicm} is supplanted by an arbitrary rational constant or even an arbitrary constant, the right-hand side of \eqref{Kicm} persistently represents a symmetry of the KdV equation for the fixed $n$-soliton solution \eqref{soln}. This observation subtly suggests that $\Phi^\alpha$, for any $\alpha$, could potentially function as a (formal) recursion operator for the KdV equation. In a parallel vein, drawing from the expressions \eqref{KBEscj} associated with the Burgers equation, one can ascertain that the application of the recursion operator $\Phi_b^i$ on the space translation symmetry $v_{ns,x}$ results exclusively in linear combinations of the soliton center translation symmetries $v_{ns,c_j}$, accompanied by the combination constants $k_j^i$ for positive $i$. Indeed, the linear admixtures of $v_{ns,c_j}$ that incorporate the combination constants $k_j^\alpha$, for $\alpha$ being any constant, continue to embody symmetries of the Burgers equation for the specific solution $v_{ns}$, thereby implying that $\Phi_b^\alpha$ acts as a formal recursion operator for any $\alpha$. To date, we have yet to uncover an explicit representation of $\Phi^\alpha$ for the KdV equation. Fortuitously, by incorporating Grassmann variables for $\alpha = 1/2$ and the variables elucidated in references \cite{Ren,Ren1} (or anyon variables delineated in \cite{anyon}) for $\alpha = 1/N$, with $N$ being any arbitrary positive integer, we can readily derive the closed, explicit form of $\Phi^\alpha$ for the Burgers equation \eqref{BE}.

Should we regard the field $v$ as a complex bosonic field within a ren space \cite{Ren,Ren1} characterized by $\{x,\ t,\ \theta\}$, then the arbitrary $\alpha$ root of $\Phi_b$, as delineated in \eqref{phib}, can be articulated as
\begin{equation}
\Phi_{r}=\Phi_b^{1/\alpha}={\cal{R}}+({\cal{R}}^{1-\alpha}v)+({\cal{R}}v)\partial_x^{-1},\quad {\cal{R}}=\partial_x^{1/\alpha}=\partial_{\theta}+\frac1{[(\alpha-1)!]_q}\theta^{\alpha-1}\partial_x, \label{psir}
\end{equation}
wherein $\alpha$ denotes a positive integer. The ren-variable $\theta$ exhibits the properties $$\theta^{\alpha}=0,\ \theta^i
\neq 0,\ 1\leq i<\alpha-1,$$
with
$$q=\exp(2{\mbox{i}}\pi \alpha^{-1}),\quad q^{\alpha}=1,$$
 and
 $$[p!]_q\equiv 1_q 2_q\cdots p_q,\ i_q\equiv 1+q+q^2+\cdots +q^{i-1}.$$
For the specific case of $\alpha=2$, the ren-variable $\theta$ is tantamount to the Grassmann variable, and the ren-symmetric derivative ${\cal{R}}$ transmutes into the well-established super derivative ${\cal{D}}$, thereby rendering the recursion operator \eqref{psir} as
\begin{equation}
\Phi_{s}=\sqrt{\Phi_b}={\cal{D}}+({\cal{D}}^{-1}v)+({\cal{D}}v)\partial_x^{-1},\quad {\cal{D}}=\sqrt{\partial_x}=\partial_{\theta}+\theta\partial_x. \label{psis}
\end{equation}
The veracity of \eqref{psir} and \eqref{psis} can be empirically validated through direct application of the Cole-Hopf (CH) transformation
\begin{equation}
v=(\ln f)_x,\ \sigma^v=\partial_x (f^{-1}\sigma^f),\ \sigma^f=f\partial_x^{-1}\sigma^v \label{fv}
\end{equation}
that mediates between the Burgers equation \eqref{BE} and the heat equation
\begin{equation}
f_{t}=f_{xx}. \label{heat}
\end{equation}
It becomes evident that ${\cal{R}}$ and ${\cal{D}}$, as defined in \eqref{psir} and \eqref{psis}, constitute the recursion operators of the heat equation \eqref{heat}. Consequently, the CH transformations of ${\cal{R}}$ and ${\cal{D}}$, namely $\Phi_r$ and $\Phi_s$, are construed as the recursion operators of the Burgers equation \eqref{BE}.
Manifestly, ${\cal{R}}f$ (${\cal{D}}f$) represents a symmetry of the heat equation. Thus, we arrive at a unified formulation for the ren-symmetric, supersymmetric, and standard integrable Burgers hierarchy,
\begin{equation}
v_{t_n}=\zeta_n {\Phi_r}^nv_x,\ n=-\alpha,\ -\alpha+1,\ \ldots,\ \infty, \label{vtn}
\end{equation}
wherein $\zeta_n$ may be a ren-number with a degree $\alpha-[n],\ [n]=(n,\ \mbox{mod}(\alpha))$, if $t_n$ is an even variable. Alternatively, $\zeta_n=1$ if $t_n$ is a ren-variable with a degree $[\partial_{t_n}]=[n]$.

In instances where $n=\alpha m$, \eqref{vtn} transitions to the standard classical Burgers hierarchy due to the identity ${\cal{R}}^{\alpha}=\Phi_b$. When $\alpha=2$, \eqref{vtn} evolves into a supersymmetric Burgers hierarchy for $n=2m-1$, given by
\begin{equation}
v_{t_{2m-1}}=\zeta_m {\Phi_s}^{2m-1}v_x,\ m=0,\ 1,\ 2,\ \ldots,\ \infty, \label{vtm}
\end{equation}
where, $\zeta_m$ might be a Grassmann constant for even time $t_{2m-1}$ or a conventional constant for odd time (for instance, $\sqrt{\partial_{\tau_m}}=\partial_{t_{2m-1}}$).
For $m=1$, \eqref{vtm} transforms into ($v={\cal{D}}\Psi$),
\begin{equation}
\Psi_{t_{1}}=\zeta_1 {\cal{D}}(\Psi_x+\Psi {\cal{D}}\Psi),\ \label{vt1}
\end{equation}
wherein $\Psi={\cal{D}}^{-1}v$ stands as a fermionic superfield, $\zeta_1$ potentially embodies a Grassmann constant for an even time $t_1$ or $\zeta_1=1$ for an odd time $t_1$ determined by $\partial_{t_1}=\sqrt{\partial_{\tau_1}}$ with even time $\tau_1$.

\section{ A symmetry conjecture}
Beyond the realm of multiple soliton solutions in integrable systems, a plethora of alternative multi-wave solution types emerges. These include, but are not limited to, multiple periodic wave solutions \cite{HuXB}, algebro-geometric solutions \cite{AG}, and multi-complexiton solutions, among others \cite{MaCom,LouCom}.
For the broader spectrum of solutions, we proffer the following conjecture:
\\
\bf Conjecture: \rm \em
For a symmetry integrable nonlinear partial differential equation system, if there is an $n$-wave solution,
\begin{equation}
u = u_{nw} = u(\xi_1,\ \xi_2,\ \ldots, \ \xi_n),\ \xi_i \equiv k_i x + \omega_i t + c_i,\ i=1,\ 2,\ \ldots,\ n, \label{Nw}
\end{equation}
where $c_i$ and $k_i$ are arbitrary constants, $\omega_i \equiv \omega_i(k_j, m_j,\ j=1, \ldots, n)$ may depend not only on the wave numbers $k_j$ but also potentially on other types of parameters such as the Riemann invariants $m_j$ associated with algebro-geometric solutions, then for the special solution \eqref{Nw}, the infinitely many $K$-symmetries and $\tau$-symmetries determined by
$K_m=\Phi^m u_x$ (with $K_m'=\Phi^m u_t$ if $u_t \neq \Phi u_x$), and  $\tau_m=\Phi^m(atu_t+bxu_x+cu)$ for $m=0,\ 1,\ 2,\ \ldots,\ \infty$ with appropriately selected constants $\{a, b, c\}$ and the recursion operator $\Phi$, are the linear combinations of the parameter translation symmetries like the center translation symmetries, $u_{c_i}=\partial_{c_i}u_{nw}$, the wave number translation symmetries, $u_{k_i}=\partial_{k_i}u_{nw}$, and any supplementary wave parameter (if extant) translation symmetries.
\rm

Should the $n$-wave solutions be identified as the $n$-soliton solutions, the conjecture delineated herein has been empirically validated for the KdV and Burgers equations. Moreover, I have meticulously examined and affirmed the veracity of this conjecture across an array of integrable systems, encompassing not only multiple soliton waves but also breathers and complexitons. These systems include, though not limited to, the modified KdV equation, the Sawada-Kortera equation, the nonlinear Schr\"odinger equation, the Boussinesq equation, the Sharma-Tasso-Olver equation, as cited in reference \cite{LouCom}, and the sine-Gordon equation.

For instance, the KdV equation \eqref{KdV} features a particular two-wave solution, known as the complexiton solution \cite{MaCom},
\begin{equation}
u_{com}=2k_1k_2\frac{(k_2^2-k_1^2)\cos(\xi_1)\sinh(\xi_2)+2k_1k_2[\sin(\xi_1)\cosh(\xi_2)-1]}
{\left[k_1\sinh(\xi_2)+k_2\cos(\xi_1)\right]^2} \label{com}
\end{equation}
with $\xi_1=k_1x-k_1(k_1^2-3 k_2^2)t+c_1$, $\xi_2=k_2x-k_2(3 k_1^2-k_2^2)t+c_2$, and the set $\{k_1,\ k_2,\ c_1,\ c_2\}$ comprising arbitrary constants. Confirming the correctness of the conjecture for this two-wave solution, $u_{com}$, is straightforward. The infinitely many $K$-symmetries and $\tau$-symmetries defined by \eqref{K} and \eqref{tau} are linear combinations of four elementary wave parameter translation invariances $u_{com,c_1}$, $u_{com,c_2}$, $u_{com,k_1}$, and $u_{com,k_2}$. In essence, only four of these symmetries, specifically $\{K_1,\ K_2,\ \tau_2,\ \tau_3\}$, are independent for the distinctive solution $u_{com}$, with all others expressible in terms of these four. The final results are encapsulated by
\begin{eqnarray}
&&K_{m\geq 3}=\frac{(-1)^{m+1/2}}{4k_1k_2}(C_{m1}K_1+C_{m2}K_2),\\
&&\tau_{m\geq 4}=\frac{(-1)^{m+1/2}}{4k_1k_2}(C_{m1}\tau_2+C_{m2}\tau_3),\
\label{KnTn}
\end{eqnarray}
where
$$C_{m1}=(k_1-\text{\rm i}k_2)^2(k_1+\text{\rm i}k_2)^{2m-4}-(k_1+\text{\rm i}k_2)^2(k_1-\text{\rm i}k_2)^{2m-4},$$
 and
$$C_{m2}=(k_1+\text{\rm i}k_2)^{2m-4}-(k_1-\text{\rm i}k_2)^{2m-4},$$
with $\text{\rm i}=\sqrt{-1}$.

\section{Solve multi-soliton solutions via generalized K-symmetries}
The utilization of Lie point symmetries stands as a cornerstone for the elucidation of similarity solutions. The intrinsic symmetries of an integrable system provide a fertile ground for the systematic extraction of a plethora of solutions, encompassing single soliton, periodic cnoidal waves, and Painlev\'e solutions. This section delineates a novel methodology for the derivation of multi-soliton solutions, predicated upon the deployment of generalized $K$-symmetries in tandem with the symmetry conjecture previously posited.
\subsection{Multi-soliton solutions of the Burgers equation via K-symmetries and symmetry conjecture.} For the The Burgers equation, denoted by \eqref{BE}, is a paradigmatic instance of a nonlinear PDE. The formulation of its general $n$-wave solutions is encapsulated within the following construction
	\begin{eqnarray}
	v=V(\xi_1,\ \xi_2,\ \ldots,\ \xi_n,\ m_1,\ m_2,\ \ldots,\ m_M)\equiv V,\ \xi_i\equiv k_ix+\omega_it+c_i,
	\label{vV}
	\end{eqnarray}
	where $\{k_i,\ \omega_i,\ c_i,\ i=1,\ 2,\ \ldots,\ n\}$ and  $\{m_j\}$ for $j = 1, 2, \ldots, M$ are arbitrary constants.
	The process of inferring the $n$-wave solutions necessitates the substitution of the aforementioned expression \eqref{vV} into the Burgers equation \eqref{BE}, yielding a pivotal condition
	\begin{eqnarray}
	\sum_{i=1}^n\left(\omega_iV_{\xi_i}-k_i\sum_{j=1}^nk_jV_{\xi_i\xi_j}+2k_iVV_{\xi_i}\right)=0.
	\label{BE1}
	\end{eqnarray}
	The incursion of the $n$-wave solution \eqref{vV} into the initial $K$-symmetries, specifically $K_1$ and $K_2$, engenders two trivial relationships, which are as follows
	\begin{eqnarray}
	K_1=\sum_{i=1}^nk_iV_{\xi_i},\ K_2=\sum_{i=1}^n\omega_iV_{\xi_i}.
	\label{K12}
	\end{eqnarray}
	The application of the $n$-wave solution assumption \eqref{vV} to the aforementioned symmetry conjecture begets an infinite series of constraints upon the $n$-wave solution, delineated by the following equations
	\begin{eqnarray}
	&K_3:& \sum_{j=1}^n a_{3j}V_{\xi_j}=\sum_{i,j}k_i(\omega_jV_{\xi_i\xi_j}+k_jV_{\xi_i}V_{\xi_j})
	+V\sum_i(\omega_i+k_iV)V_{\xi_i},  \label{bK3}\\
	&K_4:& \sum_{j=1}^n a_{4j}V_{\xi_j}=\sum_{i,j}\omega_j(\omega_iV_{\xi_i\xi_j}+2k_iV_{\xi_i}V_{\xi_j})
	+2V^2\sum_i \omega_iV_{\xi_i},  \label{bK4}\\
	&K_5:& \sum_{j=1}^n a_{5j}V_{\xi_j}=\sum_{i,j,m}k_m(\omega_i\omega_j V_{\xi_i\xi_j\xi_m} +3k_i\omega_jV_{\xi_i}V_{\xi_j\xi_m}+k_ik_jV_{\xi_i}V_{\xi_j}V_{\xi_m})+V^3\sum_i (2\omega_i+k_iV)V_{\xi_i}\nonumber\\
	&&
	+\sum_{i,j}[V(\omega_i+2k_iV)\omega_jV_{\xi_i\xi_j}+2(\omega_i\omega_j
	+2k_i\omega_jV+k_ik_jV^2)V_{\xi_i}V_{\xi_j}],  \label{bK5}\\
	&K_{i}:& \sum_{j=1}^n a_{ij}V_{\xi_j}=K_i{\big|}_{v=V,\ v_x=\sum_ik_iV_{\xi_i},\ v_{xx}=\sum_{i,j}k_ik_jV_{\xi_i\xi_j},\ \ldots,} ,\ i>5.\label{bKi}
	\end{eqnarray}
	These constraints are characterized by an array of undetermined constants $a_{ij}$, where $i \geq 3$ and $j = 1, 2, \ldots, n$.
	Given the multitude of constraints delineated by equations \eqref{bK3} through \eqref{bKi}, which are intrinsic to the $n$-wave solution \eqref{vV}, it becomes feasible to ascertain a specific solution from the variant Burgers equation \eqref{BE1} within the framework of these imposed symmetries.
	By incorporating dispersion relations
	\begin{equation}
	\omega_j=k_j^2,\ a_{3j}=k_j^3,\ a_{4j}=k_j^4,\ \ldots,\ a_{ij}=k_j^i,\ i\geq 3, \label{aij}
	\end{equation}
	for the Burgers hierarchy, the derivation of $n$-soliton solutions becomes tractable.
	The computational prowess of symbolic algebra systems, such as Maple, is harnessed to facilitate the direct computation of these solutions. By issuing the appropriate commands  `` Solve(subs(n=2,\{Eq. \eqref{BE1}, \ Eq. \eqref{bK3}\})),\ Solve(subs(n=3,\{Eq. \eqref{BE1}, \ Eq. \eqref{bK3}, \ Eq. \eqref{bK4}\})),\ Solve(subs(n=4,\{Eq. \eqref{BE1}, \ Eq. \eqref{bK3},\ Eq. \eqref{bK4},\ Eq. \eqref{bK5}\}))" in Maple, one can obtain $2$-soliton, $3$-soliton, and $4$-soliton solutions, respectively.

 \subsection{Multi-soliton solutions of the PKdV equation via K-symmetries and symmetry conjecture.}
In the rigorous mathematical treatment of the PKdV equation, as denoted by equation \eqref{pkdv}, the formulation of the general $n$-wave solutions is endowed with the following structure
	\begin{eqnarray}
	p=P(\xi_1,\ \xi_2,\ \ldots,\ \xi_n,\ m_1,\ m_2,\ \ldots,\ m_M)\equiv V,\ \xi_i\equiv k_ix+\omega_it+c_i,
	\label{pP}
	\end{eqnarray}
	where $\{k_i,\ \omega_i,\ c_i,\ i=1,\ 2,\ \ldots,\ n\}$ and $\{m_1,\ m_2,\ \ldots,\ m_M\}$ are arbitrary constants.
	\\
	Employing the $n$-wave solution assumption \eqref{pP}, the PKdV equation \eqref{pkdv} is transformed into the following form
	\begin{eqnarray}
	\sum_{i=1}^n\omega_iP_{\xi_i}=\sum_{i,j,k}k_ik_jk_mP_{\xi_i\xi_j\xi_m}+3\sum_{i,j}k_ik_jP_{\xi_i}P_{\xi_j}.
	\label{pkdv1}
	\end{eqnarray}
	The $K$-symmetries, which are infinite in number, are articulated as
	\begin{eqnarray}
	K_i=\bar{\Phi}^{i-1}p_x,\ i=1,\ 2,\ \ldots,\ \infty,\ \bar{\Phi}\equiv \partial_x^2+4p_x-2\partial_x^{-1}v_{xx}.
	\label{pKi}
	\end{eqnarray}
	Substituting the $n$-wave solution \eqref{pP} into the $K$-symmetries \eqref{pKi} yields a sequence of expressions that delineate the symmetries of the PKdV equation
	\begin{eqnarray}
	&&K_1,K_2:\ K_1=\sum_{i=1}^nk_iP_{\xi_i},\ K_2=\sum_{i=1}^n\omega_iP_{\xi_i}, \label{pK12}\\
	&&K_3:\ \sum_{j=1}^n a_{3j}P_{\xi_j}=\sum_{i,j,r}k_ik_j\omega_rP_{\xi_i\xi_j\xi_r}
	-\left(\sum_{i,j}k_ik_jP_{\xi_i\xi_j}\right)^2-2\left(\sum_{i}k_iP_{\xi_i}\right)^3
	+4\sum_{i,j}\omega_ik_jP_{\xi_i}P_{\xi_j},  \label{pK3}\\
	&&K_4:\ \sum_{j=1}^n a_{4j}P_{\xi_j}=\sum_{i,j,r}\omega_ik_j(\omega_rP_{\xi_i\xi_j\xi_r}+4k_rP_{\xi_i}P_{\xi_j}P_{\xi_r})
	+3\left(\sum_i \omega_iP_{\xi_i}\right)^2 -2\sum_{i,j,m,r,s}k_ik_jk_mk_rk_sP_{\xi_i}P_{\xi_j\xi_m}P_{\xi_r\xi_s},
	\nonumber\\
	&&\qquad +2\sum_{i,j,r,s}k_ik_jk_r(\omega_sP_{\xi_i}P_{\xi_j\xi_r\xi_s}-\omega_sP_{\xi_i\xi_j}P_{\xi_r\xi_s}
	-2k_sP_{\xi_i}P_{\xi_j}P_{\xi_r}P_{\xi_4}),  \label{pK4}\\
	&&K_5: \sum_{j=1}^n a_{5j}P_{\xi_j}=\sum_{i,j,m}\omega_i\omega_j(\omega_m V_{\xi_i\xi_j\xi_m} +18k_mV_{\xi_i}V_{\xi_j}V_{\xi_m})-6\sum_{i,j,m,r,s}k_ik_jk_mk_r\omega_sP_{\xi_i\xi_j}P_{\xi_s}P_{\xi_r\xi_m}
	\nonumber\\
	&& \qquad
	-3\sum_{i,j,r,s}k_ik_j\omega_r (\omega_sP_{\xi_i\xi_j}P_{\xi_r\xi_s}-2\omega_sV_{\xi_i}V_{\xi_j\xi_r\xi_s}
	-4k_sP_{\xi_i}P_{\xi_j}P_{\xi_r}P_{\xi_s}),
	\label{pK5}\\
	&&K_i: \sum_{j=1}^n a_{ij}P_{\xi_j}=K_i{\big|}_{p=P,\ p_x=\sum_ik_iP_{\xi_i},\ p_{xx}=\sum_{i,j}k_ik_jP_{\xi_i\xi_j},\ \ldots,} \ i>5. \label{pKi1}
	\end{eqnarray}
	Here, $a_{ij},\ i\geq3,\ j=1,\ 2,\ \ldots, n$ are constants that remain undetermined, awaiting further specification through the dispersion relations.
	\\
	Theoretically, the incorporation of the infinite constraints \eqref{pK3} to \eqref{pKi1} into the $n$-wave assumption \eqref{pP} allows for the precise determination of the $n$-wave solutions of the PKdV equation \eqref{pkdv}. For instance,, by introducing dispersion relations of the form
	\begin{equation}
	\omega_j=k_j^3,\ a_{ij}=k_j^{2i-1},\ i=3,\ 4,\ \ldots,\ \label{paij}
	\end{equation}
	one can ascertain the $n$-soliton solutions from equations \eqref{pkdv1} and \eqref{pK3} to \eqref{pKi1}.
	Specifically, invoking the Maple command ``Solve(subs(n=2,\{Eq. \eqref{pkdv1}, Eq. \eqref{pK3}, Eq. \eqref{pK4}, Eq. \eqref{pK5}\}))", facilitates the extraction of the $2$-soliton solutions of the PKdV equation \eqref{pkdv1}.

\section{Conclusions and discussions}
In the domain of integrable systems, the inexhaustible multitude of symmetries is a cornerstone that begets an equally boundless ensemble of conservation laws. This correspondence is not a mere mathematical artifact, it is the linchpin of our insight into the profound dynamics that govern these systems. The present scholarly exposition reveals that the presence of an infinite array of $K$-symmetries and $\tau$-symmetries is synonymous with the potential to manifest an infinite spectrum of $n$-wave solutions, where $n$ is arbitrary and unbounded.
The $K$-symmetries, in their infinite multitude, are tantamount to the symmetries of $n$ wave center translations, while the $\tau$-symmetries encapsulate a more intricate structure, equating to a single Galilean (or Lorentzian) symmetry, $n$ wave number translation symmetries, $n$ wave center translation symmetries, and additional symmetries pertaining to other wave parameter translations. Stated succinctly, for a fixed $n$-wave solution where $n$ is finite, only a finite subset of $K$-symmetries and $\tau$-symmetries stand as independent entities, with the remaining symmetries being expressible as linear combinations of this finite set, excluding the Galilean symmetry.
The veracity of these assertions has been meticulously established within the purview of this scholarly treatise, specifically for the $n$-soliton (or solitary wave) solutions of the KdV and Burgers equations. Moreover, for the $n$-soliton solutions, corroboration has been sought and obtained for a selection of other integrable systems.

For the nonlinear physical systems, which possess an infinite array of degrees of freedom and are articulated by PDEs, the issue of the completeness of their infinitely many symmetries presents a formidable challenge. This paper demonstrates that the purportedly infinite $K$-symmetries and $\tau$-symmetries are, in reality, incomplete. This is evidenced by their transformation into a finite set when applied to specific $n$-wave solutions. A refined perspective is introduced by suggesting the potential to identify an infinitely many novel unknown symmetries associated with particular fixed solutions. To elucidate this concept, the solitary wave solution, denoted by \eqref{p0}, within the framework of the PKdV equation \eqref{pkdv}, is examined. The complete set of these infinitely many symmetries is encapsulated by the formulation \eqref{sj}. In contrast, the traditional $K$-symmetries are delineated by the specific condition where $\lambda_i = 0$ within the broader context of \eqref{sj}.

The application of higher-order symmetries to identify precise solutions within systems of PDEs is an issue of considerable complexity. Nonlocal symmetries, in particular, have demonstrated their efficacy in the discovery of algebro-geometric solutions \cite{Cao} and in facilitating the execution of Darboux transformations \cite{LouLi,TangL}. Leveraging the symmetry conjecture introduced in Section 6 of this work,  it becomes feasible to employ higher-order local symmetries in the quest to reveal multi-wave solutions. This is especially pertinent to the elucidation of $n$-soliton solutions, which can be achieved through the introduction of dispersion relations characteristic of soliton solutions.  In the general theoretical sense, the introduction of an additional higher-order symmetry is synonymous with the potential to incorporate an additional wave (one more set of wave parameters), into the  multi-wave solutions.

For the $n$-wave solution of an integrable system, the linear combination of symmetries associated with the center translation invariance, parameterized by the combination coefficients $\sqrt[\alpha]{k_i},\ i=1,\ \ldots,\ n$ for an arbitrary exponent $\alpha$, implies that the $\alpha$-th root of the recursion operator, $\Psi \equiv \sqrt[\alpha]{\Phi}$, retains the properties of a recursion operator. For linearizable C-integrable systems, such as the Burgers and Liouville equations, the $\alpha$-th root of the recursion operator can be explicitly derived through the utilization of the ren-number and the ren-symmetric derivative(s). Notably, for $\alpha$ equals $2$, the ren-number and ren-symmetric derivative correspond precisely to the Grassmann variable and the super derivative, respectively. Employing the $\alpha$-th root of the recursion operators, it is conceivable to subsume various types of ren-symmetric integrable hierarchies, supersymmetric integrable hierarchies, and classical integrable hierarchies under a unified framework. This unification is a profound concept that has been substantiated within the context of C-integrable hierarchies.

Furthermore, it is acknowledged that integrable systems may harbor additional special solutions characterized by further free parameters beyond the wave center and wave number parameters. For instance, in the case of algebro-geometric solutions of genus $n$, there exist ancillary parameters akin to Riemann parameters that are tied to the periods of sub-waves.
The methodologies, conclusions, and conjectures presented herein, when extended to the broader class of integrable systems, may well engender novel methodologies for tackling these systems. These potential advancements will be the subject of future scholarly discourse.

\section*{Acknowledgments}
 {\noindent \small
 The work was sponsored by the National Natural Science Foundations of China (Nos.12235007, 11975131).
 The author is indebt to thank Profs. Q. P. Liu, X. B. Hu, R. X. Yao, X. Z. Hao, M. Jia, X. G. Geng, Z. N. Zhu, B. F. Feng, J. P. Wang and J. Y. Wang for their helpful discussions.}

\end{CJK*}
\end{document}